\pdfoutput=1
\documentclass[pdftex,aps,prl,superscriptaddress,twocolumn]{revtex4}
\usepackage{amsmath}
\usepackage{amssymb}
\usepackage{amsfonts}
\usepackage{amsthm}
\usepackage[pdftex]{graphicx}
\usepackage{subfigure}
\usepackage{bm}
\usepackage{hyperref}
\usepackage{rotating}

\renewcommand{\vec}[1]{\mathbf{#1}}

\renewcommand{\Im}{\operatorname{Im}}

\def\beas{\begin{eqnarray*}}
\def\eeas{\end{eqnarray*}}
\def\bea{\begin{eqnarray}}
\def\eea{\end{eqnarray}}
\def\be{\begin{equation}}
\def\ee{\end{equation}}

\newcommand{\bpm}{\begin{pmatrix}}
\newcommand{\epm}{\end{pmatrix}}
\newcommand{\bmm}{\begin{matrix}}
\newcommand{\emm}{\end{matrix}}

\newcommand{\citeasnoun}[1]{Ref.~\onlinecite{#1}}

\begin{document}

\title{Casimir repulsion between metallic objects in vacuum}

\author{Michael Levin}
\affiliation{Department of Physics, Harvard University, 
Cambridge MA 02138}

\author{Alexander P. McCauley}
\author{Alejandro W. Rodriguez}
\author{M. T. Homer Reid}
\affiliation{Department of Physics, Massachusetts Institute of 
Technology, Cambridge MA 02139}

\author{Steven G. Johnson}
\affiliation{Department of Mathematics, Massachusetts Institute of 
Technology, Cambridge MA 02139}

%\date{\today}

\begin{abstract}
We give an example of a geometry in which two metallic objects in 
vacuum experience a repulsive Casimir force. The geometry consists of 
an elongated metal particle centered above a metal plate with a 
hole. We prove that this geometry has a repulsive regime using a 
symmetry argument and confirm it with numerical calculations for both 
perfect and realistic metals. 
%The repulsion is unambiguous as it occurs
%when the two objects lie on opposite sides of an imaginary separating 
%plane. 
The system does not support stable levitation, as the particle is
unstable to displacements away from the symmetry axis.
%The particle is unstable with respect to displacements 
%away from the symmetry axis, so that all equilibria are unstable, 
%consistent with a recent theorem.
\end{abstract}


\maketitle

% ----------------------------------------------------------------

\begin{figure}[t]
\centerline{
\includegraphics[width=1.0\columnwidth]{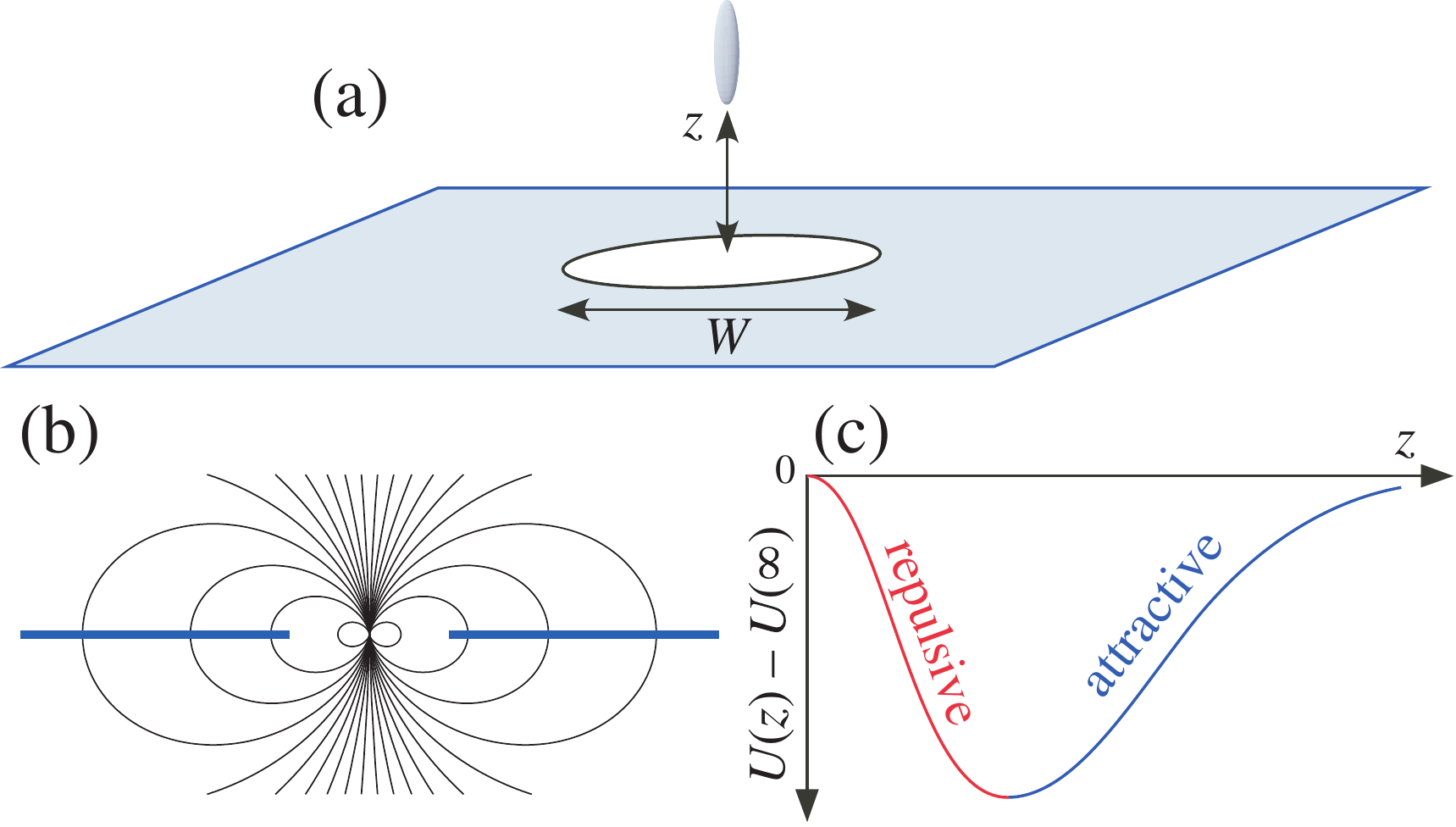}
}
\caption{(color online) (a) Schematic geometry achieving Casimir
  repulsion: an elongated metal particle above a thin metal plate 
  with a hole. The idealized version is the limit of an infinitesimal
  particle polarizable only in the $z$ direction. (b) At $z=0$,
  vacuum-dipole field lines are perpendicular to the plate by
  symmetry, and so dipole fluctuations are unaffected by the plate
  (for \emph{any} $\omega$, shown here for $\omega=0$). (c) Schematic
  particle--plate interaction energy $U(z)-U(\infty)$: zero at $z=0$
  and at $z\to\infty$, and attractive for $z \gg W$, so there must be
  Casimir repulsion (negative slope) close to the plate.}
\label{dipfield}
\end{figure}

\emph{Introduction}: The Casimir force between two parallel metal 
plates in vacuum is always attractive. A longstanding question is 
whether this is generally true for metallic/dielectric objects in 
vacuum, or whether the sign of the force can be changed by geometry 
alone. More precisely, can the force between non-interleaved 
metallic/dielectric bodies in vacuum---that is, bodies that lie on 
opposite sides of an imaginary separating plane---ever be repulsive?
In this paper, we answer this question in the affirmative by showing 
that a small elongated metal particle centered above a thin metal plate 
with a hole, depicted in Fig.~\ref{dipfield}(a), is repelled from the 
plate in vacuum when the particle is close to the plate. The particle 
is unstable to displacements away from the symmetry axis, so that
the system does not support stable levitation, consistent with 
the theorem of \citeasnoun{Rahi10:PRL}. We establish our result using a symmetry 
argument for an idealized case and by brute-force numerical calculations 
for more realistic geometries and materials. We also show that this 
geometry is closely related to an unusual electrostatic system in which a
neutral metallic object \emph{repels} an electric dipole (in fact, one 
can even obtain electrostatic repulsion for the case of a point 
charge~\cite{LevinJo10}). Anisotropic particles are
essential here; a spherical particle above a perforated plate is always
attracted, although non-monotonic effects in an isotropic case have been
suggested for the null-energy condition rather than the Casimir
energy~\cite{GrahamOl05}.

Casimir repulsion is known to be impossible for 
1d/multilayer~\cite{LambrechtJe97} or  
mirror-symmetric~\cite{KennethKl06} metallic/dielectric geometries in
vacuum. Interleaved ``zipper'' geometries can combine attractive
interactions to yield a separating ``repulsive''
force~\cite{RodriguezJo08:PRA}, but the sign of the force is
ambiguous in such geometries. (In contrast, in this paper the 
objects lie on opposite sides of a separating $z=0^+$ plane and the 
interaction is unambiguously repulsive.)  Repulsive forces also arise 
for fluid-separated geometries~\cite{Munday09} or 
magnetic~\cite{Boyer74,Kenneth02} or 
magnetoelectric materials~\cite{Zhao09,GrushinCo10}. A repulsive Casimir 
pressure was predicted within a hollow metallic sphere~\cite{Boyer68, 
Milton78}, but this is controversial as it does not correspond to a 
rigid-body motion, is intrinsically cutoff-dependent~\cite{Jaffe03}, 
and the repulsion disappears if the sphere is cut in 
half~\cite{KennethKl06}. Another proposal is to use ``metamaterials'' 
formed of metals and dielectrics arranged into complex 
microstructures~\cite{Henkel05, Leonhardt07, Rosa08:PRL,Zhao09}.  However,
no specific metamaterial geometries that exhibit Casimir repulsion have 
been proposed, and the theoretical result~\cite{Rahi10:PRL} 
indicates that repulsion in the metamaterial limit (separations $\gg$ 
microstructure) is impossible for parallel plate geometries. 

\emph{Symmetry argument}: We begin by establishing repulsion in an 
idealized geometry: an infinitesimal particle centered 
above an infinitesimally thin perfect-metal plate with a hole. We 
assume the particle is electrically polarizable only in the $z$~direction 
and is not magnetically polarizable at all (the limit of an infinitesimal 
metallic ``needle'') and the plate lies in 
the $z = 0$ plane [Fig.~\ref{dipfield}(a)]. The Casimir(--Polder)
energy for such a particle at position $\vec{x}$ is given 
by~\cite{McLachlan63}
\begin{equation}
U(\vec{x}) =  
-\frac{1}{2\pi}\int_0^\infty \alpha_{zz}(i\xi) \langle E_z(\vec{x}) 
E_z(\vec{x}) \rangle_{i\xi} d\xi.
\label{casenergy}
\end{equation}
Here $\alpha_{zz}$ is the electric polarizability of the particle in the 
$z$~direction and $\langle E_z E_z
\rangle_{i\xi}$ is the mean-square $z$ component of the electric-field
fluctuations at position $\vec{x}$ and imaginary frequency 
$\omega = i\xi$. This expectation value is evaluated in a geometry 
without the particle (i.e. a geometry consisting of only the 
perforated plate in vacuum). Conventionally, it is 
renormalized by subtracting the (formally infinite) mean-square 
fluctuations in vacuum. One way to compute the expectation value is to
note that it is related to a classical electromagnetic Green's 
function via the fluctuation-dissipation theorem. More specifically, 
$\langle E_z(\vec{x}) E_z(\vec{x'})\rangle_{\omega}$ is proportional 
to the electric field $E_z(\vec{x'})e^{-i\omega t}$ produced by an 
oscillating $z$-directed electric dipole 
$\vec{p} = p_z \hat{\vec{z}}e^{-i\omega t}$ at position $\vec{x}$.

The key idea for establishing repulsion is simple: we find a 
point $\vec{x}$ \emph{such that the classical field of an 
oscillating $z$-directed electric dipole at $\vec{x}$ is unaffected by the 
presence of the metallic plate with a hole}. It then follows that 
$U(\vec{x}) = U(\infty)$, implying that the energy $U$ must vary 
non-monotonically between $\vec{x}$ and $\infty$ and hence must be 
repulsive at some intermediate points. While in most geometries no 
such $\vec{x}$ exists, in the perforated plate geometry this condition 
is achieved by symmetry at $\vec{x} = 0$. If a $z$-directed electric dipole is 
placed at $z=0$ in the hole, then the electric-field lines of the 
dipole in vacuum are \emph{already} perpendicular to the plate by 
symmetry, as illustrated in Fig.~\ref{dipfield}(b); thus, the 
\emph{vacuum} dipole field solves Maxwell's equations with the correct
boundary conditions in the presence of the plate, and 
$U(z=0) = U(\infty)$.  Note that this is true by symmetry at 
\emph{every} frequency $\omega$ (real or imaginary), because the 
dipole moment $\vec{p}$ at $z=0$ is antisymmetric with respect to the 
$z=0$ mirror plane. Intuitively, the basic point is that the electric dipole
fluctuations of the particle do not couple to the plate at all when
$z = 0$.

For large $z$---that is, $z$ much larger than the hole diameter 
$W$---the presence of the hole in the plate is negligible, and we 
must have the usual attractive Casimir--Polder interaction. So, as 
schematically depicted in Fig.~\ref{dipfield}(c), we expect the 
interaction energy $U(z) - U(\infty)$ to be zero at $z=0$, decrease to
negative values for small $z > 0$ (leading to a repulsive force) then 
increase to zero for large $z$ (leading to an attractive force). If 
the hole is circular, then by symmetry the force is purely in the $z$ 
direction and the point of minimum $U$ is an equilibrium position, 
stable under $z$ perturbations; however, both the theorem 
of~\citeasnoun{Rahi10:PRL} and explicit calculations show that this 
equilibrium point is unstable under lateral ($xy$) perturbations of
the particle position. In fact, numerical calculations (not shown) indicate 
that the particle is unstable to lateral perturbations and tilting 
at all separations $z$.

\begin{figure}[tb]
\centerline{
\includegraphics[width=0.8\columnwidth]{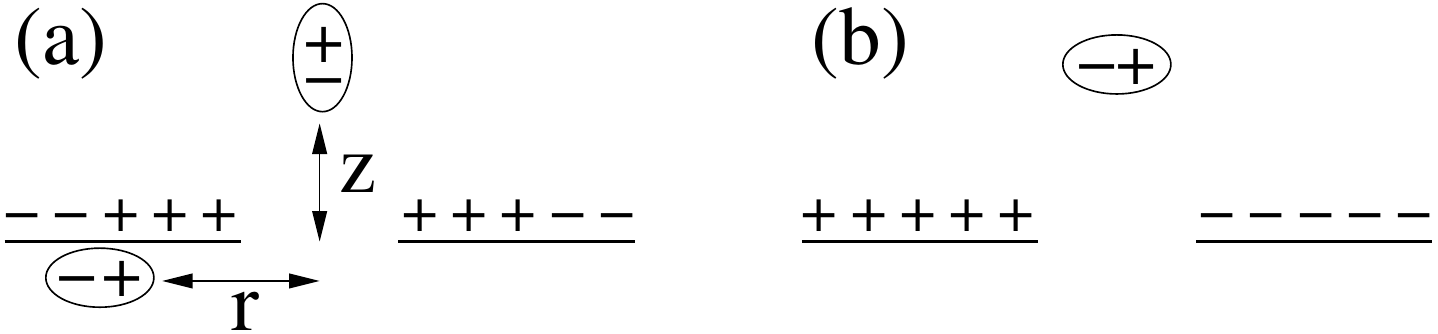}
}
\caption{ (a) Schematic electrostatic interaction of a dipole with a
  neutral perforated plate (side view), depicting the charge
  density $\sigma$ induced on the plate.  Since $\sigma$ is positive
  for small $r$ and negative for large $r$, $\sigma$ can be
  constructed out of a superposition of dipoles in the $z=0$ plane,
  oriented radially inward about the $z$ axis. A simple calculation
  then shows that the interaction is repulsive for small $z$. (b) In 
  contrast, a dipole oriented parallel to the plate is always attracted,
  as one can see from the induced charge density shown above.}
\label{esrep}
\end{figure}

\emph{Electrostatics}:
Strictly speaking, this symmetry argument only shows that the force
must be repulsive at \emph{some} $z \neq 0$: $U$ could 
conceivably have multiple oscillations. To definitively rule out this 
possibility, we rely on the explicit numerical calculations described 
below. However, on an intuitive level, the basic behavior of
the force can be understood from electrostatic considerations.

\begin{figure}[tb]
\centerline{
\includegraphics[width=1.0\columnwidth]{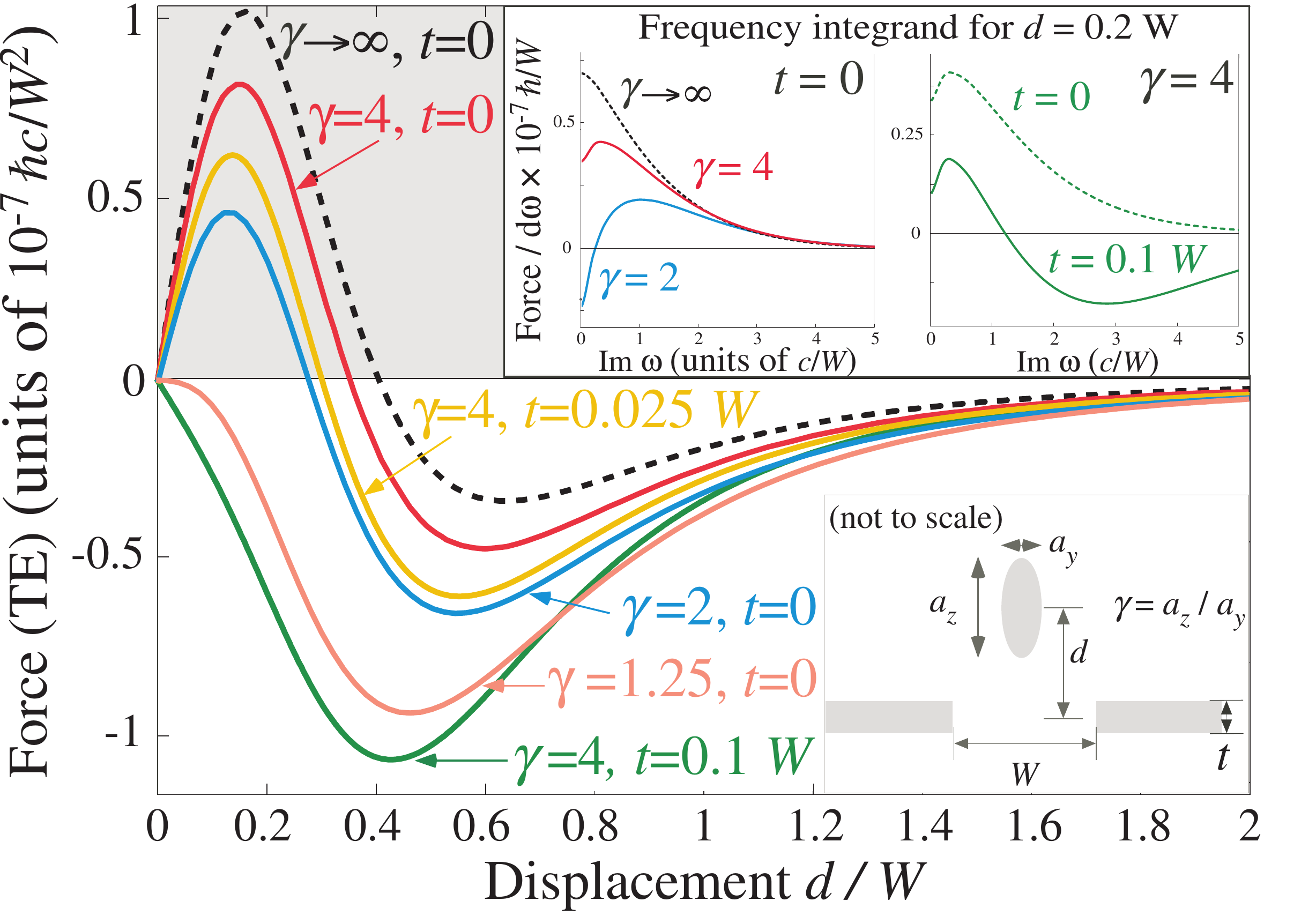}
}
\caption{(color online) Exact (BEM) 2d Casimir force for
  ellipse--line geometry (lower inset) with perfect metals and 
  TE polarization (in-plane electric field); positive force (shaded) 
  is repulsive. The effects of both particle width $a_y$ and line 
  thickness $t$ are shown, for fixed $a_z = 0.002W$.  As the 
  ellipticity $\gamma = a_z/a_y$ decreases or $t$ increases, the 
  repulsive force diminishes.  Upper-left inset: frequency-resolved 
  force $F(\Im \omega)$ at fixed separation $d=0.2 W$ and $t=0^+$: as 
  $\gamma$ decreases, attractive contributions arise from small 
  $\Im \omega$.  Upper-right inset: $F(\Im \omega)$ at fixed $d=0.2W$ 
  and $\gamma = 4$: as $t$ increases, attractive contributions arise
  from large $\Im \omega$.}
\label{data-2d}
\end{figure}

To see this, let us focus on the $\omega = i\xi = 0$ contribution to 
the Casimir energy (\ref{casenergy}); we expect the contribution from 
nonzero imaginary frequencies to be qualitatively similar (though
this expectation can sometimes be violated, as in the inset of 
Fig.~\ref{data-2d}). The 
$\omega=0$ contribution is proportional to the electrostatic energy 
of a $z$-directed electric dipole in the presence of a neutral metal plate. By 
the same arguments as above, such an electrostatic dipole must be 
repelled from the plate for some $z > 0$. To see this explicitly, 
suppose there is a static dipole at some position $(0,0,z)$, and 
consider the induced charges on the plate. In the limit where the 
plate is infinitesimally thin, we can combine the charges on the two 
sides of the plate into a single surface charge density $\sigma$. On a
qualitative level, we expect this total charge density to be of the 
form shown in Fig.~\ref{esrep}(a), with $\sigma$ positive for small $r$ 
and negative for large $r$. In particular, $\sigma$ can be constructed
out of a superposition of dipoles in the $z=0$ plane, oriented 
radially inward about the $z$ axis. A simple calculation shows that 
vertical force on a dipole at $(0,0,z)$ from a horizontal dipole at 
distance $r$ from the $z$ axis is repulsive if $r > 2z$ and attractive
if $r < 2z$. Thus, if the hole is circular with diameter $W$ and 
$z < W/4$, all the dipoles will exert a repulsive force and the total 
force is necessarily repulsive. On the other hand, when $z \gg W$, 
most of the dipoles will exert an attractive force, so the total force
is attractive. In contrast, a dipole oriented 
\emph{parallel} to the plate is always attracted, as one can see 
from the induced charge density schematically shown in 
Fig.~\ref{esrep}(b). This explains why an elongated shape is
necessary for the repulsive effect (see Fig.~\ref{data-2d}): dipole
fluctuations parallel to the plate give rise to an attractive Casimir
force.  

We can confirm this picture by solving this electrostatics problem 
exactly in the two-dimensional (2d) case, where the metal 
plate with a hole is replaced by a metal line with a gap of width 
$W$. Assuming a 2d Coulomb force $F(r_{12}) = q_1 q_2/r_{12}$, and a 
$z$-directed dipole moment $p_z$, we find
\begin{equation}
U_\mathrm{electrostatic}(z) = -p_z^2 \cdot 
\frac{2z^2}{(W^2+4z^2)^2}.
\label{exact2d}
\end{equation} 
The force is indeed repulsive for small $z$, with a sign change 
occurring at $z = W/2$. A similar calculation for a $y$-directed 
dipole yields a uniformly attractive force.

The repulsion in the $z$-directed case is quite unusual, even in 
electrostatics: in almost all
cases, the electrostatic interaction between an electric dipole and a neutral
metal object is attractive, not repulsive. Indeed, on an intuitive 
level, it seems almost inevitable that a dipole will induce a dipole moment
in the metal object oriented so that the force is attractive. More rigorously, 
one can prove that this interaction 
is attractive in several different limits. For example, if a dipole 
is very far away ($z \rightarrow \infty$) or very close to the surface 
of a metal object, the interaction is always attractive. One can also 
prove that the force is attractive if the metal object is replaced by a
dielectric material with a permittivity 
$\epsilon/\epsilon_0 = 1 + \delta$ where $0 < \delta
\ll 1$, using a perturbative expansion in $\delta$. Clearly, a special
geometry is necessary to obtain a repulsive force in electrostatics,
and arguably in Casimir interactions as well by extension to
$\omega\neq 0$ along the imaginary frequency axis (see concluding
remarks below).

\begin{figure}[tb]
\centerline{
\includegraphics[width=1.0\columnwidth]{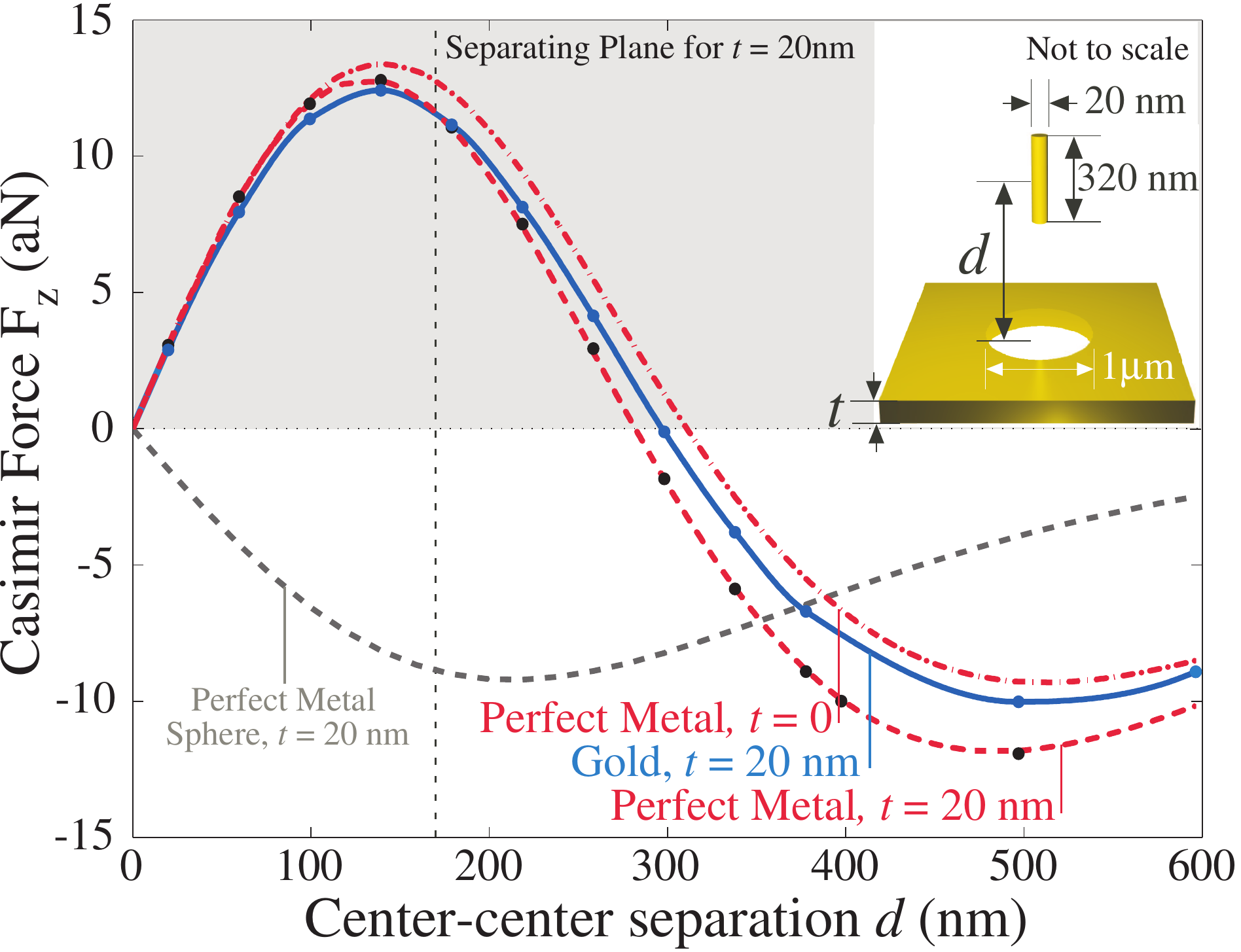}
}
\caption{(color online) Exact Casimir force for 
  cylinder--plate geometry (inset) for perfect metals (computed with BEM) 
  and gold (computed with FDTD); positive (shaded) force is repulsive.  
  For $d \lesssim 300$~nm and $d > 170$~nm (vertical dashed line), the 
  force is unambiguously repulsive as the cylinder is entirely above
  the plate. In contrast, a perfect-metal sphere (diameter $60$~nm) 
  is always attracted to the plate.}
\label{force3d}
\end{figure}

\emph{Numerical demonstration}: Moving beyond the idealized geometry, 
we expect the repulsion to be robust under small perturbations, such as
finite particle size, plate thickness, and permittivity. This 
expectation is validated by the explicit calculations described below.
We utilize two recent numerical methods, evaluating
the Casimir force at zero temperature. First, 
we use a finite-difference time-domain (FDTD) 
technique that computes the Casimir stress tensor via the Green's
function~\cite{RodriguezMc09:PRA, McCauleyRo10:PRA}, with a
free-software implementation~\cite{Oskooi10:Meep}. Second, we use a
boundary-element method (BEM) that can solve either for the stress
tensor or directly for the Casimir energy/force via a path-integral
expression~\cite{ReidRo09}. 

Figure~\ref{force3d} shows the Casimir force for a finite-size
cylindrical metal particle ($20\times 320$~nm)
above a finite-thickness ($t=20$~nm) plate with a circular hole of
diameter $1\,\mu$m, considering both perfect metals and 
finite-permittivity gold, along with the force on a
perfect-metal sphere (diameter $60$~nm) for comparison. The perfect metal
results were computed with BEM; the others were  
computed by our FDTD technique in cylindrical coordinates, with 
the gold permittivity described by 
$\epsilon(\omega = i\xi) = 1 + \omega_p^2/\xi^2$, where $\omega_p =
1.37\times 10^{16}$ rad/sec (the omission of the loss term, which is
convenient for FDTD~\cite{RodriguezMc09:PRA}, does not significantly
affect our results). The force on the sphere is always attractive, 
while the force on the cylindrical particle is repulsive for a 
center--center separation $d \lesssim 300$~nm. Because of the finite 
sizes, when $d < 170$~nm the tip of the particle intrudes into the 
hole. However, there is a range of about $130$~nm for $d > 170$~nm where 
the force is unambiguously repulsive: the two 
objects lie on opposite sides of an imaginary separating plane. 
Similar behavior is seen for an infinitesimally thick ($t=0$) plate
(Figure~\ref{force3d}). 
Somewhat surprisingly, the finite-permittivity gold exhibits a 
\emph{stronger} repulsive force than the perfect-metal case of the 
same geometry; this is explained below as a consequence of the finite 
thickness of the plate. 

In order to better understand the dependence on geometry, we use
BEM to explore the parameter space of a simplified 2d
($yz$) version of the problem: a metal elliptical particle above a 
metal line with a gap of width $W$. We compute the Casimir force
in this setup for perfect metals and 2d electromagnetism, with the
standard convention that the electric field is in the plane, as in a 
``TE'' mode. (This system is equivalent to a scalar field
with Neumann boundary conditions on the two objects). In 
Figure~\ref{data-2d}, we explore how both the 
ellipticity $\gamma = a_z/a_y$ of the particle and the line thickness 
$t$ affect the force, for a fixed width $W$ and particle length 
$a_z = 0.002W$.  As $\gamma \rightarrow 1$, the elliptical particle 
becomes increasing circular, and repulsion diminishes due to the 
attractive force associated with dipole fluctuations in the $y$ 
direction. We find that the repulsive force disappears for 
$\gamma \lesssim 1.25$, when $t =0$.  
Similarly, as $t$ becomes larger, one can no longer make the 
approximation that the metal line does not affect the field of a 
$z$-directed dipole at $d=0$, and the repulsive effect disappears by 
$t \approx 0.1 W$ for $\gamma = 4$.

Further insight can be gained from the contribution of each imaginary 
frequency $\omega = i\xi$ to the force for a fixed particle--line
separation $d=0.2W$ (roughly maximum repulsion).  The upper-left inset
to Fig.~\ref{data-2d} shows that as $\gamma$ decreases, the attractive
contributions first appear at \emph{small} $\xi$, eventually
making the overall force attractive.  The right half of the inset
shows that, in contrast, a nonzero $t$ gives rise to attractive force
contributions at \emph{large} $\xi$. This may explain
the larger repulsive force of real gold compared to perfect
metal in Fig.~\ref{force3d}: the finite skin depth of gold
cuts off the large-$\xi$ contributions, reducing the attractive
effects of the finite plate thickness (which, in this case, dominate
the attractive effects of the finite particle size and ellipticity).

\emph{Concluding remarks}: In this paper, we have shown that the sign 
of the Casimir force in vacuum can be changed by geometry alone, 
without ``cheating'' by interleaving the bodies as 
in~\citeasnoun{RodriguezJo08:PRA}. Consistent with 
~\citeasnoun{Rahi10:PRL}, the geometry described here does not support 
stable levitation since the particle is unstable with respect to 
lateral ($xy$) translation and tilting (as shown by additional 3d BEM 
calculations). As for the question of experimental realizations, we leave
this to future work, though we note that one approach would be to anchor 
the particle to a substrate plate made of a low permittivity material using a 
pillar made of the same kind of material. Assuming a periodic array of 
such pillars and a complementary array of holes with a unit cell of area 
$10 (\mu\text{m})^2$, and estimating the 
force per unit cell as the maximum repulsive force calculated in 
Fig.~\ref{force3d}, one obtains a repulsive pressure of about $10^{-6}$
Pa. This is 2 or 3 orders of magnitude smaller than typical experimental 
sensitivities \cite{KlimMo09}, but the repulsion could be increased 
further by shrinking or optimizing the geometry. 

This geometry was motivated by the electrostatic analogue shown in 
Fig. \ref{esrep}, where a qualitatively similar effect is observed.
Previously, another interesting non-monotonic Casimir effect was also
seen to have an electrostatic analogue~\cite{RahiRo07}. 
Mathematically, the mostly non-oscillatory exponential decay of the 
Casimir-force contributions for imaginary frequencies $\omega=i\xi$ 
tends to make the total force qualitatively similar to the $\xi\to0^+$ 
contribution (in fact, this similarity becomes an exact proportionality 
in the unretarded, van der Waals limit). This suggests that one 
approach for discovering ``interesting'' geometric Casimir effects is to 
first find an interesting electrostatic interaction, and then seek an 
analogous Casimir system. 
%For example, one could imagine engineering a perforated plate 
%in the shape of a \emph{curved} equipotential surface of a dipole, 
%rather than a flat plane. Repulsion in the electrostatic plate-dipole 
%problem is then guaranteed by the same argument as above. However, the
%physics of the Casimir analogue---consisting of the curved perforated 
%plate and an elongated particle---is more complex. In particular, the
%plate will not be orthogonal to the vacuum dipole field lines at 
%nonzero frequencies, leading to the possibility of interesting 
%``geometric dispersion'' effects where different frequencies give
%competing attractive/repulsive contributions to the Casimir force.

This work was supported in part by the Harvard Society of Fellows,
the US DOE grant DE-FG02-97ER25308, and the DARPA contract 
N66001-09-1-2070-DOD.

\bibliographystyle{apsrev}

\end{document}